\documentclass[11pt,twoside]{article}

\usepackage{asp2006}
\usepackage{epsf}
\usepackage{lscape}

\begin{document}
\title{3-D Dynamics of Interactions between Stellar Winds and the 
Interstellar Medium as Seen by AKARI and Spitzer}   

\author{%
Toshiya Ueta$^1$, 
Hideyuki Izumiura$^2$,
Issei Yamamura$^3$, 
Robert E.\ Stencel$^1$,
Yoshikazu Nakada$^{4,5}$,
Mikako    Matsuura$^{6}$,
Yoshifusa Ita$^{3,6}$,
Toshihiko Tanab\'{e}$^{4}$,
Hinako    Fukushi$^{4}$,
Noriyuki  Matsunaga$^{6}$,
Hiroyuki  Mito$^{5}$, and 
Angela K.\ Speck$^7$}   
\affil{%
$^1$ Dept.\ of Physics \& Astronomy, University of Denver, USA\\
$^2$ Okayama Astrophysical Observatory, NAOJ\\
$^3$ Institute of Space and Astronautical Science, JAXA \\
$^4$ Institute of Astronomy, Schooof Science, University of Tokyo\\
$^5$ Kiso Observatory, Institute of Astronomy, University of Tokyo\\
$^6$ National Astronomical Observatory of Japan\\
$^7$ Dept.\ of Physics \& Astronomy, University of Missouri, USA}

\begin{abstract} 
Recent far-infrared mapping of mass-losing stars by the
 {\sl AKARI Infrared Astronomy Satellite} and {\sl Spitzer Space
 Telescope} have suggested that far-infrared bow shock structures are
 probably ubiquitous around these mass-losing stars, especially
 when these stars have high proper motion.
Higher spatial resolution data of such far-infrared bow shocks now allow
 detailed fitting to yield the orientation of the bow shock cone with
 respect to the heliocentric space motion vector of the central star,
 using the analytical solution for these bow shocks under the assumption
 of momentum conservation across a physically thin interface between the
 stellar winds and interstellar medium (ISM).
This fitting analysis of the observed bow shock structure would enable
 determination of the ambient ISM flow vector, founding a new technique
 to probe the 3-D ISM dynamics that are local to these interacting
 systems.  
In this review, we will demonstrate this new technique for three
 particular cases, Betelgeuse, R Hydrae, and R Cassiopeiae.
\end{abstract}



\section{Interactions between Stellar Winds and the ISM}

Most stars lose their surface matter into the surrounding interstellar
space in one way or another via mass loss (e.g.\
\citealt{willson00,kp00}). 
Such mass loss processes intimately relate these mass-losing stars to
the interstellar medium (ISM), which these stars enrich with ashes of
nuclear burning taking place inside these stars.
Indeed, mass loss is the critical process that one must
consider when it comes to the mass budget of the ISM (e.g.\
\citealt{sedlmayr94,mm09}).

When we think of such ISM enrichment by stellar mass loss ejecta,
however, we often na\"{\i}vely expect that stellar ejecta will
eventually meet with the ISM and get merged into it without giving
serious thought about the process of the stellar ejecta merging
with the ISM.
On the first order approximation, the wind-ISM merger occurs where 
the ram pressure of the stellar wind is balanced by the pressure of 
the ambient ISM: a shock is expected to form if the
wind velocity is greater than the ambient sound speed ($v_{\rm w} >
v_{\rm c}$) and a pile-up of the ejecta is anticipated otherwise
($v_{\rm w} < v_{\rm c}$).  
Either way, a density enhancement ensues at the interface between the
stellar wind and ISM.
Such density enhancements can be detected if sufficient amount of
radiation is induced within the wind-ISM interface regions.
Since the wind-ISM interface often occurs surrounding the mass-losing
star, we typically detect such density enhancements in the form of 
concentric arcs around the central star.

Such arc-shaped circumstellar structures had already been found around
luminous OB stars and Wolf-Rayet and red supergiant stars
\citep{stencel88,vbm88} in the far-infrared (far-IR) All-Sky maps
obtained by the {\sl Infrared Astronomy Satellite} ({\sl IRAS\/}).   
Far-IR emission from these arcs is probably mainly due to thermal 
emission of the cold dust component of the density enhancement at the
wind-ISM interface whose shock-heated temperature peaks at far-IR. 
There may be contributions from low-excitation atomic lines such as 
[\ion{O}{I}] $63\micron$, [\ion{O}{I}] $145\micron$ and
[\ion{C}{II}] $158\micron$ at these wavelengths. 
However, the exact emission mechanism of these far-IR bow shocks
remains unclear. 
While an attempt to identify their spectroscopic nature is currently
on-going with the {\sl Spitzer Space Telescope} ({\sl Spitzer\/};
\citealt{wer04}), this is beyond the scope of the present paper.  
Below, we will restrict our discussion to concentrate on the structure
of these stellar wind-ISM bow shocks and what we can learn from it.  

\section{3-D Motions of the Star, Shock, and the ISM}

{\sl IRAS} discoveries of these bow shock structures at the wind-ISM
interface were certainly astounding.
Nevertheless, the spatial resolution of the {\sl IRAS} observations
(typically $2\arcmin$ to $5\arcmin$) was not sufficient enough to allow detailed
structural analyses of the far-IR shock surfaces.  
To pursue this avenue of research, we had to wait until the far-IR
renaissance of this decade with the coming of {\it Spitzer} and the {\it
AKARI Infrared Astronomy Satellite} ({\sl AKARI\/}: \citealt{murakami07}),
which permit sub-arcminute spatial resolution in the far-IR.

\begin{figure}[!t]
\begin{center}
   \resizebox{0.9\hsize}{!}{
     \includegraphics*{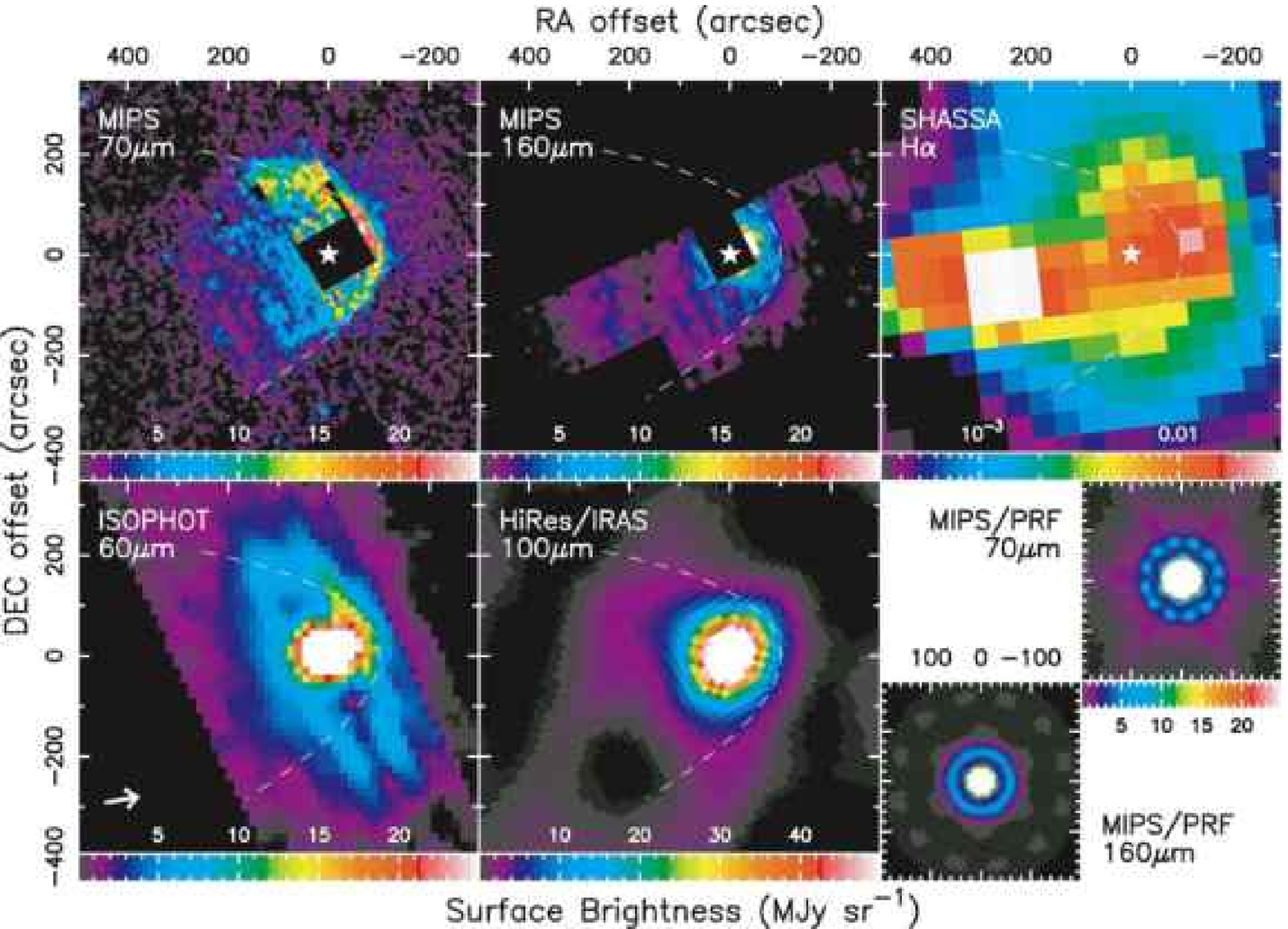}
   }
\end{center}
\caption{%
{\sl Spitzer} maps of R Hya at 70 and $160\micron$ (top left and middle, 
 respectively; also shown are the point-response function at these
 wavelengths at the bottom right) along with an {\sl Infrared Space
 observatory} map at $60\micron$ (bottom left; \citealt{hashimoto97})
 and an {\sl IRAS} map at $100\micron$ (bottom middle; HiRes-processed)
 as well as a SHASSA H$\alpha$ map (top right; \citealt{gaustad01}).
 Tick marks indicate angular offsets in arcseconds with respect
 to the position of the star.
 Surface brightness in MJy sr$^{-1}$ is indicated by the scale 
 at the bottom of each panel. 
 A parabolic curve, which roughly represents the shape of the bow shock,
 is displayed by the dashed lines.  
The arrow at the bottom left corner shows the direction of the proper
 motion of the star, $(\mu_{\alpha}, \mu_{\delta})=(-57.68, 12.86)$
 mas yr$^{-1}$ \citep{leeuwen07}.  Reproduced from
 \citet{ueta06}.}\label{rhya}    
\end{figure}

\begin{figure}[!t]
\begin{center}
   \resizebox{0.9\hsize}{!}{
     \includegraphics*{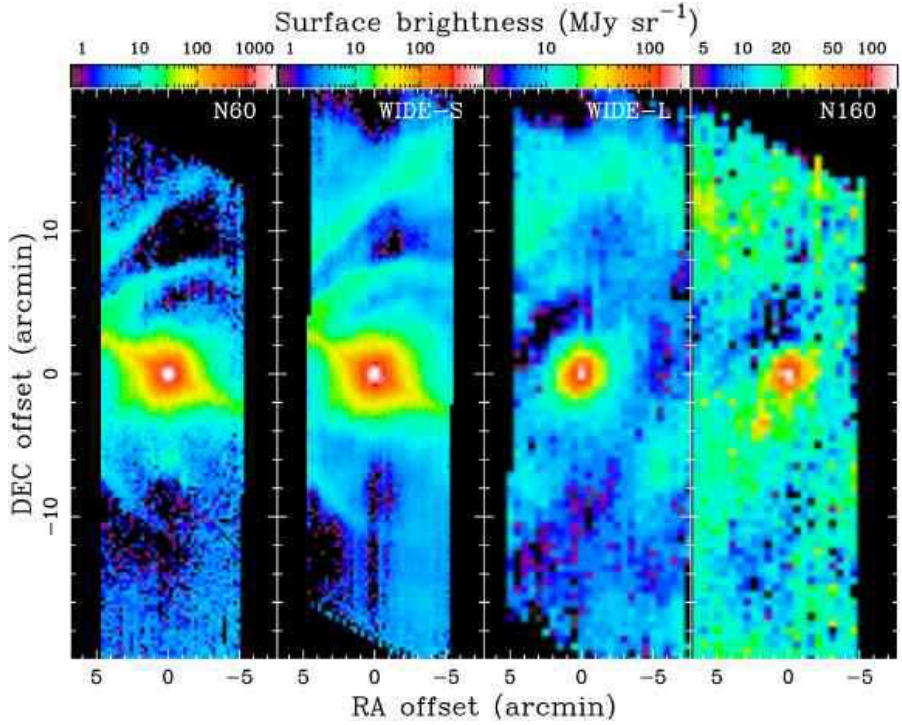}
   }
\end{center}
\caption{%
AKARI/FIS maps of Betelgeuse in the SW bands - N60 ($65\micron$) and
 WIDE-S ($90\micron$) at $15\arcsec$ pixel$^{-1}$ scale - and in the LW
 bands - WIDE-L ($140\micron$) and N160 ($160\micron$) at
 $30\arcsec$ pixel$^{-1}$ scale - from left to right, respectively.  
 Tick marks indicate angular offsets in arcminutes with respect to the
 position of the star. 
Surface brightness in MJy sr$^{-1}$ is indicated by the scale at the top
 of each panel. 
The revised {\sl Hipparcos} proper motion is $(\mu_{\alpha},
 \mu_{\delta})=(27.54, 11.30)$ mas yr$^{-1}$ \citep{leeuwen07}.
Reproduced from \citet{ueta08}.}\label{aori}  
\end{figure}

This far-IR renaissance in context of the present study began with a
{\sl Spitzer} discovery of a stellar wind bow shock arc around R Hydrae
(R Hya), a
Mira-type evolved star 
(\citealt{ueta06}, Figure \ref{rhya}), followed by a revisit to
Betelgeuse, one of the first {\sl IRAS\/} discoveries of a far-IR
stellar wind bow shock, using the Far-IR Surveyor instrument (FIS; 
\citealt{kawada07}) aboard {\sl AKARI} (\citealt{ueta08},
Figure \ref{aori}).  
New far-IR images of these sources clearly reveal the characteristic
arc-shaped structure of a bow shock at the interface between stellar
winds and the ISM at sub-arcminute spatial resolution.
For these bow shocks to occur, the mass-losing central star has to be
moving {\sl relative to} the ISM that is local to these objects.
Indeed, these stars exhibit proper motion that is consistent with the
orientation of the bow shocks (i.e.\ toward the apex of the bow).

Meanwhile, theoretical studies have been done both analytically
\citep{wilkin96,wilkin00} and numerically (e.g.\
\citealt{ml91,dgani96,bk98,wareing07}) to understand the structure of
these bow shocks.
Remarkably, if one assumes momentum conservation across a physically
thin (i.e.\ radiative-cooling dominating) shock layer, one can express
the shape of the bow shock analytically as a function of the latitudinal
angle $\theta$ measured from the direction of the bow apex with respect
to the position of the central star as follows \citep{wilkin96}: 
\begin{eqnarray}\label{bow}
 R(\theta) = R_0 \frac{\sqrt{3(1-\theta\cot\theta})}{\sin\theta} 
\end{eqnarray} 
where $R_0$ is the {\sl stand-off distance} between the star and bow
apex defined as 
\begin{eqnarray}\label{standoff}
 R_0 =\sqrt{\frac{\dot{M}v_{\rm w}}{4\pi\rho_{\rm ISM} v_{*}^2}}
\end{eqnarray}
for which $\dot{M}$ is the rate of mass loss,
$v_{\rm w}$ is an isotropic stellar wind velocity, 
$\rho_{\rm ISM}$ is the density of the ambient ISM, and 
$v_{*}$ is the space velocity of the star.

The observed shape of the bow shock is really a projection of the bow
shock cone (eqn.\ref{bow}) onto the plane of the sky \citep{wilkin96}.
For instance, while the R Hya bow appears very close to an edge-on bow
shock cone (Figure \ref{rhya}) the Betelgeuse bow looks rather circular
(Figure \ref{aori}), suggesting an inclination angle far from edge-on.
At far-IR, the surface brightness is proportional to the dust column
density along the line of sight.
If the bow shock is physically thin we can expect that the column
density becomes the highest where the bow shock layer intersects with
the plane of sky including the central star, based on dust radiative
transfer calculations (e.g.\ \citealt{ueta03}). 
Hence, one can confidently determine the orientation of the bow shock
cone by fitting the apparent bow shock shape with the Wilkin bow shock 
cone equation (eqn.\ref{bow}), taking into account the inclination angle
of the bow.
This {\sl Wilkin solution fitting}, therefore, would yield the
heliocentric orientation of the bow shock cone with respect to the observer. 
This orientation of the bow represents a relative motion between the
mass-losing central star and the ISM local to the star. 
At the same time, the central star's heliocentric space motion can be
defined observationally via measurements of its proper-motion and radial
velocity.
Naturally, the probability of detecting a bow shock is increased if the
star has a large space motion.

Here, we have two vectors of motion concerning a mass-losing star and
its ambient ISM. 
One is the heliocentric relative motion of the star with respect to the
ambient ISM and the other is the star's heliocentric space motion.
These two vectors are not necessarily the same because the ambient ISM
can move/flow into a certain direction irrespective of the star's space
motion. 
Therefore, if one compares the result of the Wilkin solution fitting of
the bow shock shape (i.e.\ the heliocentric relative motion of the
mass-losing central star with the ambient ISM) and the observed space
motion of the mass-losing central star (i.e.\ the heliocentric space
motion of the star), one can deduce the heliocentric flow vector of the
ambient ISM in principle.

\section{Wilkin Solution Fitting}

Now, we would like to review the potential of this investigation into the
heliocentric flow vector of the ambient ISM, which is local to a
mass-losing moving star, based on the Wilkin solution fitting
\citep{wilkin96} to the apparent bow shock shape for the following cases
of three particular stars. 

\subsection{Betelgeuse (Red Supergiant)}

Using the {\sl AKARI}/FIS image at the most sensitive WIDE-S
($90\micron$) band (second from left in Figure \ref{aori}), the best-fit
of the Wilkin solution fitting yields 
the inclination angle of the bow shock cone with respect to the plane of
the sky ($\theta_{\rm incl}$) of $\pm (56^{\circ}\pm4^{\circ})$,
position angle of $55^{\circ}\pm2^{\circ}$ (east of north), and
stand-off distance of $4\farcm8\pm0\farcm1$.
The leading $\pm$ sign of $\theta_{\rm incl}$ indicates the unresolvable
degeneracy of the fitting, i.e., the bow shock cone points either away
from or to us but the fitting is unable to determine which. 
Nevertheless, the Wilkin fitting defines a space motion vector of the
star {\sl relative to} the ambient ISM (albeit the degeneracy), which we
define as ${\mathbf v}_{* \rm ISM}$.

Meanwhile, using the most recent, multi-epoch proper-motion study of
Betelgeuse using VLA data, \citet{harper08} derived $(v_{\alpha},
v_{\delta}) = (23.3, 8.9)$ km s$^{-1}$ at 197pc.
With the average radial velocity of the star ($v_{\rm rad}$) at
$20.4\pm0.4$ km s$^{-1}$ (pointed away from us), these numbers yield another 
space motion vector of the star having $\theta_{\rm incl} = 40^{\circ}$
and the position angle of $69^{\circ}$.
This space motion vector is of course {\sl relative to} the Sun (i.e.,
the heliocentric correction applied), which we denote as ${\mathbf v}_{* \odot}$. 
Using these two space motion vectors of the star in two distinct frames,
we can then derive the heliocentric flow vector of the ambient ISM,
${\mathbf v}_{\rm ISM \odot}$, by
\begin{eqnarray}\label{vism}
 {\mathbf v}_{\rm ISM \odot} = {\mathbf v}_{* \odot} - {\mathbf v}_{* \rm ISM}.
\end{eqnarray}

\begin{figure}[!t]
\begin{center}
   \resizebox{0.6\hsize}{!}{
     \includegraphics*{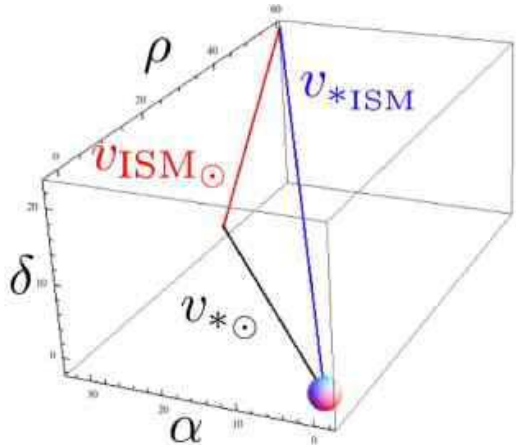}
   }
\end{center}
\caption{%
3-D relationship among the heliocentric space motion of Betelgeuse
 (${\mathbf v}_{*\odot}$), heliocentric flow of the ambient ISM
 (${\mathbf v}_{\rm ISM\odot}$), and
space motion of Betelgeuse relative to the ambient ISM (${\mathbf v}_{*
 \rm ISM}$). The sphere at the lower right corner indicates the position
 of Betelgeuse at the origin of the equatorial coordinates $(\alpha,
 \delta, \rho)$.
}\label{aori3d}   
\end{figure}

Before we proceed any further, however, we have to break the degeneracy
of ${\mathbf v}_{* \rm ISM}$. 
Since Betelgeuse is in the vicinity of the Orion Nebula Complex (ONC),
we have relatively complete information about its local environment
(e.g.\ \citet{odell01}).   
One of the bases for the present study is the presumed existence of the
ISM flow local to the mass-losing moving star.
Therefore, there must exist a source of this ISM flow, and  
the star forming regions in the ONC are most likely.
If we assume there is an overall isotropic outflow from the ONC,
Betelgeuse's proper motion suggests that it has been traversing 
across this ISM flow in front of the ONC (see Figure 4
of \citealt{harper08}).
If this is the case, the apex of the bow shock cone around Betelgeuse
should point {\sl away from us}, i.e., $\theta_{\rm incl} =
-(56^{\circ}\pm4^{\circ})$.  
Thus, the degeneracy is lifted and relevant quantities are only
dependent on the ambient ISM density, $\rho_{\rm ISM}$.

Using $R_0 = 0.3$pc \citep{ueta08} and $\dot{M} = 3 \times 10^{-6}$
M$_\odot$ yr$^{-1}$ as derived by \citet{harper08} at the best-estimated
distance of 197pc to Betelgeuse and $v_{\rm w} = 17$ km s$^{-1}$
\citep{bernat79}, Betelgeuse's space motion with respect to the ambient
ISM turns out to be
\begin{eqnarray}
 {\mathbf v}_{* \rm ISM} 
=
\left(
\begin{array}{c}
v_{\rm rad} \\
v_{\alpha} \\ 
v_{\delta} 
\end{array}
\right)
=
\frac{1}{\sqrt{n_{\rm ISM}}}
\left(
\begin{array}{c}
32.7 \\ 
18.1 \\ 
12.7 
\end{array}
\right)
=
\left(
\begin{array}{c}
59.7 \\ 
33.0 \\ 
23.2 
\end{array}
\right)
~~\mbox{(km s$^{-1}$).}
\end{eqnarray}
At the last step, we adopt the ISM number density in front of the ONC,
$n_{\rm ISM} = 0.3$ cm$^{-3}$ \citep{frisch90}.  
Therefore, the heliocentric ambient ISM flow at Betelgeuse is
\begin{eqnarray}
 {\mathbf v}_{\rm ISM \odot} 
=
{\mathbf v}_{* \odot} 
-  
{\mathbf v}_{* \rm ISM} 
=
\left(
\begin{array}{r}
20.7 \\ 
23.3 \\ 
8.9 
\end{array}
\right)
-
\left(
\begin{array}{c}
59.7 \\ 
33.0 \\ 
23.2 
\end{array}
\right)
=
\left(
\begin{array}{r}
-39.1 \\ 
-9.7 \\ 
-14.2 
\end{array}
\right)
~~\mbox{(km s$^{-1}$)}
\end{eqnarray}
with
$v_{*\odot} = 32.4$ km s$^{-1}$,
$v_{*\rm ISM} = 72.1$ km s$^{-1}$, and
$v_{\rm ISM \odot} = 42.7$ km s$^{-1}$.
3-D relationship among these three vectors is shown in 
Figure \ref{aori3d}.
Hence, Betelgeuse is traversing across the ambient ISM flow (of
42.7 km s$^{-1}$) at the heliocentric space velocity of 32.4 km s$^{-1}$,
resulting in the space velocity relative to the ambient ISM
at 72.1km s$^{-1}$.
In the frame of stellar winds, therefore, the flow velocity of the
ambient ISM is $89.1$ km s$^{-1}$, which can induce a strong shock.

Betelgeuse is located about 200 pc away from the ONC.
At 89.1 km s$^{-1}$ the crossing time is roughly 2.2 Myr, which is
consistent with the crossing time of Betelgeuse in front of the ONC.
Thus, the assumption of an outflow from the ONC being responsible for
the stellar wind-ISM bow shock of Betelgeuse is indeed plausible in
retrospect. 

\subsection{R Hydrae (AGB Star)}

Using the {\sl Spitzer} $70\micron$ map (top left in Figure \ref{rhya}),
the best-fit of the Wilkin solution fitting yields 
$\theta_{\rm incl} = \pm (59^{\circ}\pm4^{\circ})$ and
the position angle of $-62^{\circ}\pm3^{\circ}$, 
defining ${\mathbf v}_{* \rm ISM}$.
The stand-off distance of $1\farcm6\pm0\farcm1$ translates to 0.1pc
at the adopted distance of 165pc \citep{zijlstra02}.
The updated {\sl Hipparcos} proper motion measurements by
\citet{leeuwen07} give $(v_{\alpha}, v_{\delta}) = (-45.1, 10.1)$ km s$^{-1}$ at 
165pc.
With $v_{\rm rad} = -10.4$ km s$^{-1}$ (pointed to us), another space
motion vector has $\theta_{\rm incl} = 13^{\circ}$ and the position
angle of $-77^{\circ}$.

\begin{figure}[!b]
\begin{center}
   \resizebox{1.0\hsize}{!}{
     \includegraphics*{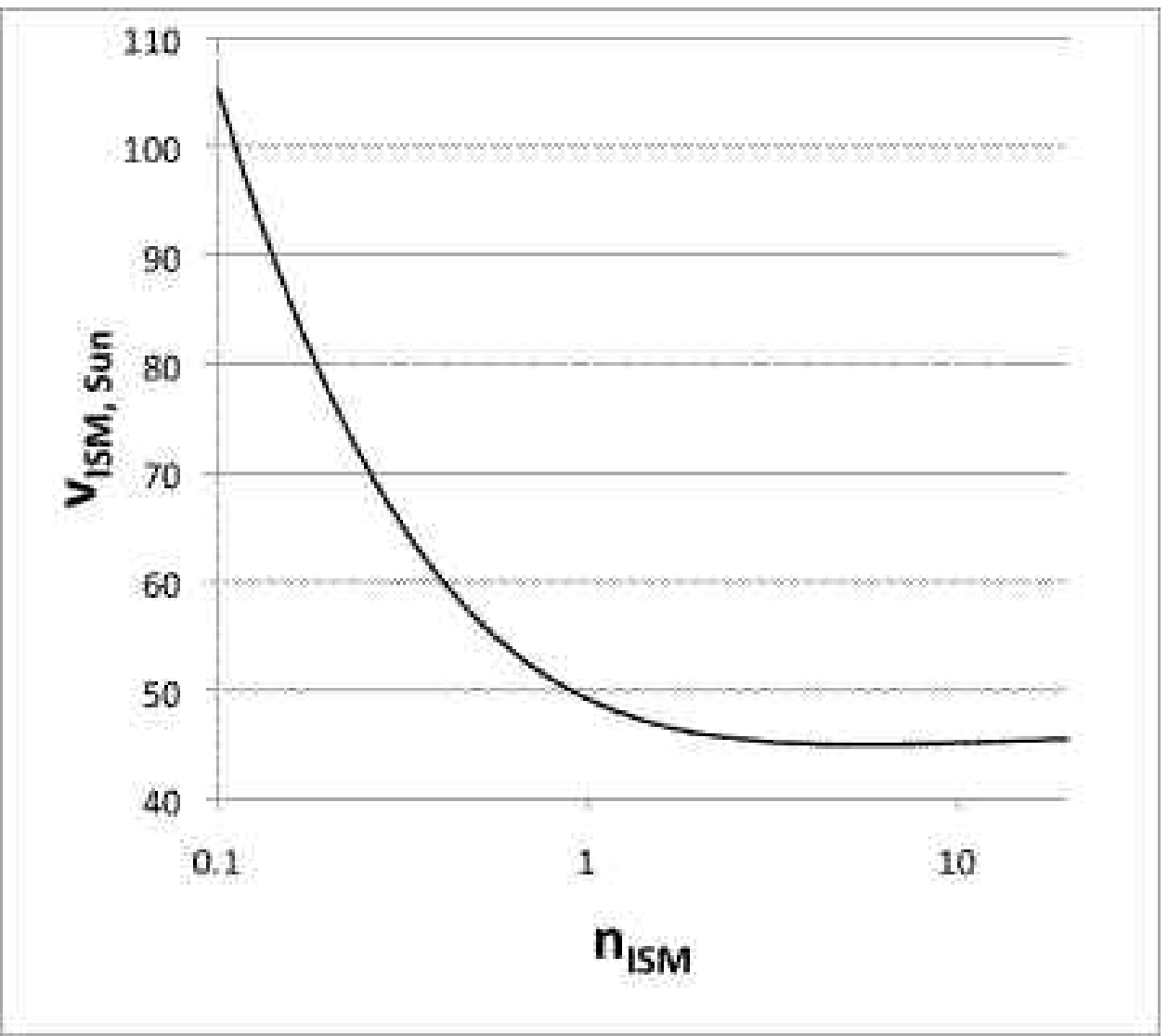}
     \includegraphics*{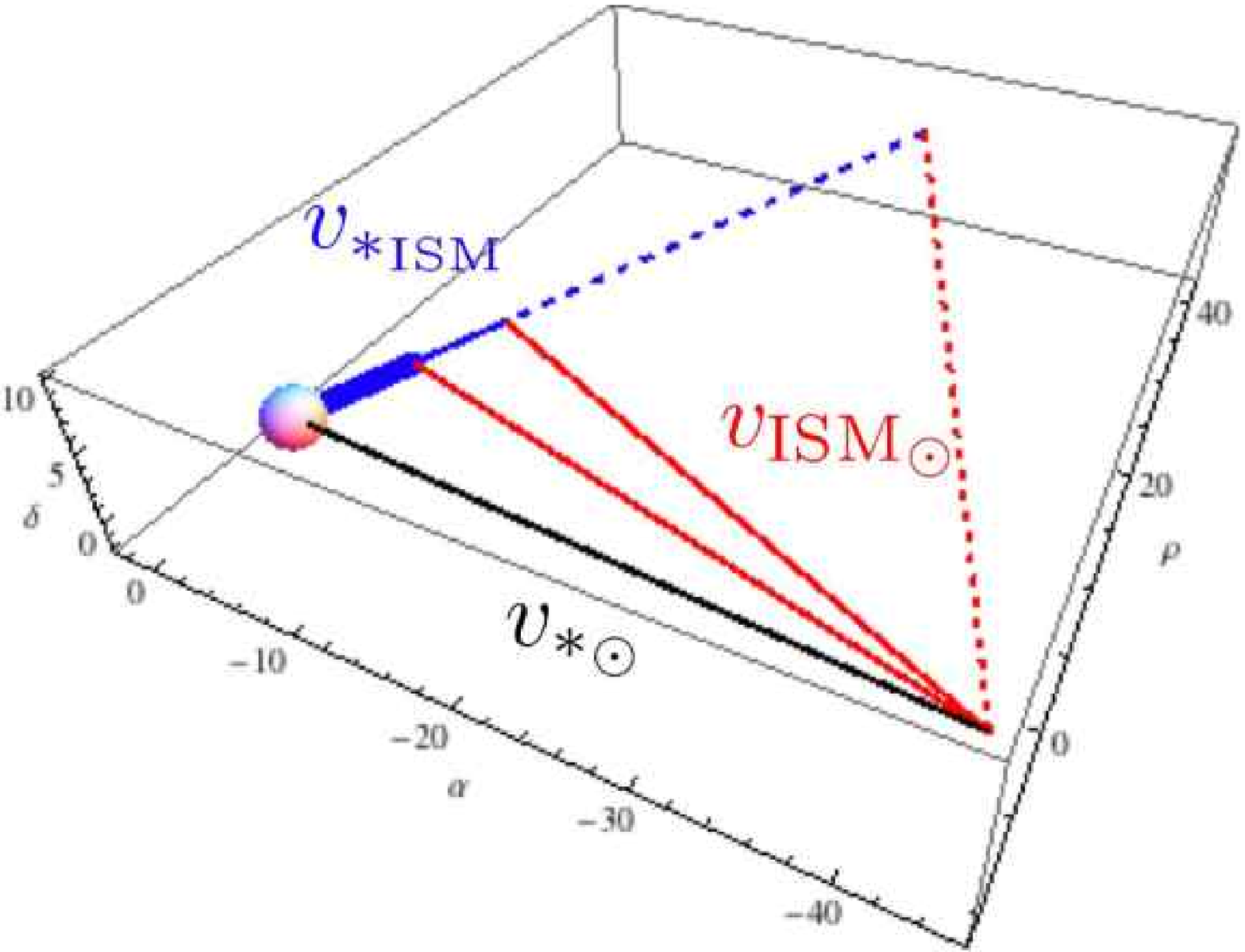}
   }
\end{center}
\caption{%
[Left] The ${\mathbf v}_{\rm ISM\odot}$-$n_{\rm ISM}$ plot for R Hya
 showing  
 behavior of ${\mathbf v}_{\rm ISM\odot}$ as a function of $n_{\rm ISM}$.
[Right] 3-D relationship among 
${\mathbf v}_{*\odot}$ 
${\mathbf v}_{\rm ISM\odot}$
and
${\mathbf v}_{* \rm ISM}$
for R Hya and its ambient ISM.
The sphere at the lower left corner indicates the position
 of R Hya at the origin of the equatorial coordinates $(\alpha,
 \delta, \rho)$.
Depending on the choice of the unknown $n_{\rm ISM}$ involved in
 ${\mathbf v}_{*\rm ISM}$, the solution to ${\mathbf v}_{\rm ISM\odot}$
 differs.  Displayed are 
(1) $n_{\rm ISM} = 20$ cm$^{-3}$ (thick solid line, ${\mathbf v}_{*\rm
 ISM} = 7.8$ km s$^{-1}$),  
(2) $n_{\rm ISM} = 5$ cm$^{-3}$ (solid line, ${\mathbf v}_{*\rm ISM} =
 14.2$ km s$^{-1}$), and 
(3) $n_{\rm ISM} = 0.5$ cm$^{-3}$ (dashed line, ${\mathbf v}_{*\rm ISM} =
 49.0$ km s$^{-1}$). 
}\label{rhya3d}   
\end{figure}

Recent CO observations have revealed a lopsided emission profile,
indicating that there is a larger amount of receding matter than
approaching matter \citep{teyssier06}. 
In context of a bow shock cone, this means that the cone is pointed away
from us (hence more of the receding component).
With this the degeneracy of ${\mathbf v}_{* \rm ISM}$ is removed and we derive
\begin{eqnarray}
  {\mathbf v}_{\rm ISM \odot} 
=
\left(
\begin{array}{c}
v_{\rm rad} \\
v_{\alpha} \\ 
v_{\delta} 
\end{array}
\right)
=
\left(
\begin{array}{r}
-10.4 \\
-45.1 \\ 
10.1 
\end{array}
\right)
-
\frac{1}{\sqrt{n_{\rm ISM}}}
\left(
\begin{array}{r}
29.9 \\ 
-17.5 \\ 
3.9 
\end{array}
\right)
~~(\mbox{km s$^{-1}$})
\end{eqnarray}
as a function of the remaining unknown, $n_{\rm ISM}$.
Unfortunately there is no definite observational diagnostic for the
value of $n_{\rm ISM}$.

However, the ${\mathbf v}_{\rm ISM\odot}$-$n_{\rm ISM}$ plot (Figure
\ref{rhya3d}, left panel) shows that the minimum value of $v_{\rm
ISM\odot}$ is $45.0$ km s$^{-1}$ when $n_{\rm ISM} = 6$ cm$^{-3}$ and that 
$v_{\rm ISM\odot}$ remains roughly $\sim 45.0$ km s$^{-1}$ for larger $n_{\rm
ISM}$ but increases for smaller $n_{\rm ISM}$.   
3-D relationship among these three vectors is shown for three different
values of $n_{\rm ISM}$ (Figure \ref{rhya3d}, right panel).  
The space velocity of R Hya relative to the ambient ISM is
$34.9/\sqrt{n_{\rm ISM}}$ km s$^{-1}$, and hence, the smaller $n_{\rm ISM}$
becomes the larger $v_{*\rm ISM}$ becomes.
In the frame of stellar winds ($v_{\rm w} = 10$ km s$^{-1}$, \citealt{knapp98}),
therefore, the flow velocity of the ambient ISM with respect to R Hya is
$10+34.9/\sqrt{n_{\rm ISM}}$ km s$^{-1}$, which can reach $\sim 
50$ km s$^{-1}$ for $n_{\rm ISM} < 0.8$ cm$^{-3}$.
Typically lower $n_{\rm ISM}$ values are preferred for high galactic
latitude objects such as R Hya.
Thus, the $v_{\rm ISM\odot}$ value (and hence, the ISM flow velocity
relative to R Hya) can be even higher. 
The next step in this line of research is to establish shock diagnostics
for low density, low velocity shocks and obtain the shock velocity in
order to fix $n_{\rm ISM}$.
This will present a new technique to probe local ISM, complementing
previous work such as \citet{lalle03}.

\subsection{R Cassiopeiae (AGB Star)}

Far-IR images of R Cassiopeiae (R Cas) show an elliptically elongated
shell around the central star that is located off-center of the shell,
but the shell does not show any obvious signature of a bow shock
\citep{ueta09}. 
However, the wind crossing time for the shell is much longer than the
time it took for the star to move to the current off-center position
from the center of the shell at the velocity known from proper motion
measurements \citep{leeuwen07}. 
Thus, we speculate that the shell shape is maintained by the
interactions between the stellar winds and ambient ISM and that the
absence of an apparent bow shock is due to its inclination angle.

\begin{figure}[!b]
\begin{center}
   \resizebox{0.6\hsize}{!}{
     \includegraphics*{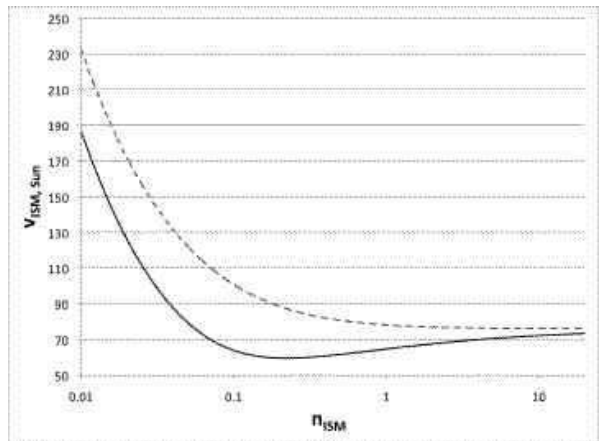}
   }
\end{center}
\caption{%
The ${\mathbf v}_{\rm ISM\odot}$-$n_{\rm ISM}$ plot for R Cas showing 
 behavior of ${\mathbf v}_{\rm ISM\odot}$ as a function of $n_{\rm ISM}$.
The solid line is the case where the bow shock cone points to us and the
 dashed line is the case where the bow shock cone points away from us.
}\label{rcasnism}   
\end{figure}

The best-fit of the Wilkin solution fitting yields 
$\theta_{\rm incl} = \pm (68^{\circ}\pm2^{\circ})$ and 
the position angle of $74^{\circ}\pm2^{\circ}$, 
defining ${\mathbf v}_{*\rm ISM}$.
The stand-off distance of $1\farcm4\pm0\farcm1$ translates to 0.1pc
at the adopted VLBI-measured distance of 176pc \citep{vlemmings05}.
The updated {\sl Hipparcos} proper motion measurements by
\citet{leeuwen07} give $(v_{\alpha}, v_{\delta}) = (71.4, 14.6)$ km s$^{-1}$ at 
176pc.
With $v_{\rm rad} = -22.9$ km s$^{-1}$ (pointed to us), another space
motion vector has $\theta_{\rm incl} = 17.5^{\circ}$ and the position
angle of $78.4^{\circ}$.
Keeping the degeneracy of ${\mathbf v}_{* \rm ISM}$ we derive
\begin{eqnarray}
  {\mathbf v}_{\rm ISM \odot} 
=
\left(
\begin{array}{c}
v_{\rm rad} \\
v_{\alpha} \\ 
v_{\delta} 
\end{array}
\right)
=
\left(
\begin{array}{r}
-22.9 \\
71.4 \\ 
14.6 
\end{array}
\right)
-
\frac{1}{\sqrt{n_{\rm ISM}}}
\left(
\begin{array}{r}
\pm20.9 \\ 
7.9 \\ 
2.3 
\end{array}
\right)
~~(\mbox{km s$^{-1}$})
\end{eqnarray}
as a function of the remaining unknown, $n_{\rm ISM}$.
The ${\mathbf v}_{\rm ISM\odot}$-$n_{\rm ISM}$ plot in Figure
\ref{rcasnism} shows that $v_{\rm ISM\odot}$ reaches the minimum of 59.5
km s$^{-1}$ at $n_{\rm ISM} = 0.22$ 
cm$^{-3}$ when the bow shock cone points to us or of 76.2 km s$^{-1}$ at
$n_{\rm ISM} = 18$ cm$^{-3}$ when the bow shock cone points away from
us.
In the frame of stellar winds at 12 km s$^{-1}$ \citep{knapp98},
the space velocity of R Cas relative to the ambient ISM is
$12+22.5/\sqrt{n_{\rm ISM}}$ km s$^{-1}$, which can reach 60 km s$^{-1}$ if the bow
shock cone is pointed to us and $n_{\rm ISM} = 0.22$ cm$^{-3}$.
For low galactic latitude objects such as R Cas, relatively higher
$n_{\rm ISM}$ values are expected.
Thus, we may expect a value close to the lower limit of 60 km s$^{-1}$.

\section{Summary}

When a bow shock structure is observed around a mass-losing star, the
central star's proper motion is often consistent with the direction to
the apex of the bow. 
To the first order, this gives an intuitive picture of the mass-losing
star moving in the otherwise stationary ambient ISM.
Higher spatial resolution far-IR images of such stellar wind bow shocks
now allow detailed fitting of the structure given the analytic shape
derived by \citet{wilkin96}.
The Wilkin solution fitting, performed for three particular cases, has 
demonstrated that the heliocentric space velocity vector of a
mass-losing star and 
the orientation of the bow shock cone do not necessarily align with each
other, suggesting the existence of a flow of the ambient ISM local to
the star.

Since the Wilkin solution fitting is inherently degenerate and other
independent observations are needed to break this degeneracy (i.e.\
determination the orientation of the bow shock cone with respect to the
plane of the sky) and to derive the density of the ambient ISM, the
analysis is not self-sufficient.
Nevertheless, three cases we have shown here unequivocally indicate that   
flows of the ambient ISM are rather prevalent and high proper-motion
stars are often traversing such ambient ISM flows, generating bow shocks
as a result. 
Therefore, stellar wind-ISM bow shocks are potentially excellent probes
of the 3-D ISM dynamics local to these shock structures.

Recent far-IR imaging surveys of mass-losing stars such as {\sl
AKARI\/}'s own MLHES Mission Program (PI: I.\ Yamamura) and its
extension for {\sl Spitzer\/} ({\sl Spitzer}-MLHES, PI: T.\ Ueta) have
identified a number of sources that seem to possess bow shock
structures. 
Using these data sets, one can additionally learn the stellar wind-ISM 
interactions besides the mass loss histories (from the undisturbed
parts of the extended dust shells) that these investigations are
originally designed to do.
With {\sl AKARI}'s All-Sky Survey data, there will be even more of these
far-IR bow shocks associated with high proper-motion stars.
Therefore, it appears that large-scale investigations of far-IR bow
shocks in context of probing the 3-D ISM dynamics should be pursued at
least for the solar-neighborhood and we intend to follow up this new
avenue of research in near future with existing {\sl AKARI} and {\sl
Spitzer} data as well as data from near-future opportunities such as
{\sl Herschel Space Observatory} and {\sl Stratospheric Observatory for
Infrared Astronomy}. 
 
\vspace*{\fill}

\acknowledgements 

This research is based on observations with {\sl AKARI\/}, a JAXA project
with the participation of ESA, and the {\sl Spitzer Space Telescope},
which is operated by JPL/Caltech under a contract with NASA.  
Support for this work was provided by University of Denver, ISAS/JAXA,
and NASA via JPL/Caltech. 


\end{document}